\documentclass[10pt]{article}
\usepackage{subfigure}
\usepackage{graphicx}
\usepackage{algorithm}
\usepackage{algorithmic}
\usepackage{setspace}
\input{epsf}
\newcounter{thedefc}
\normalsize
\title{\bf A New Benchmark For Evaluation Of  Graph-Theoretic Algorithms\footnote{This work performed under the auspices of the U.S. Department of Energy by Lawrence Livermore National Laboratory under Contract DE-AC52-07NA27344.
A technical report of this work is also available in LLNL-TR-422519.}}
\author{Andy B. Yoo$^\dagger$ \and Yang Liu$^\dagger$ \and Sheila Vaidya$^\dagger$ \and Stephen Poole$^\ddagger$ \and \\
\begin{tabular}{l}
$^\dagger$Lawrence Livermore National Laboratory, Livermore, CA 94551 \\
$^\ddagger$Oak Ridge National Laboratories, Oak Ridge, TN 37831 \\
\end{tabular}}
\date{}

\begin{document}
\maketitle
\thispagestyle{empty}
\begin{abstract}
We propose a new graph-theoretic benchmark in this paper.
The benchmark is developed to address shortcomings of an existing widely-used
graph benchmark. We thoroughly studied a large number of traditional and contemporary graph algorithms
reported in the literature to have clear understanding of  their algorithmic and run-time 
characteristics. Based on this study, we designed a suite of kernels, each of which represents a 
specific class of graph algorithms. The kernels are designed to capture the typical run-time 
behavior of target algorithms accurately, while limiting computational and spatial overhead
to ensure its computation finishes in reasonable time.
We expect that the developed benchmark will serve as a much needed tool for evaluating different 
architectures and programming models to run graph algorithms.
\end{abstract}

\section{Introduction}
\label{introduction}
Graph algorithms have become an increasingly important and widely-used tool in a wide range of
emerging disciplines such as web mining, computational biology, social network analysis, and text analysis.
Typically, input to these algorithms are the graphs that consist of typed vertices and typed edges that
represent the relationships between the vertices. The vertices and edges of the graphs are often
associated with some attributes.

These graphs, which belong to a class of graphs called scale-free graphs (or networks)~\cite{barabasi1999:emergence},
have very complex structures.
Furthermore, the graphs that are usually formed by fusing fragmental information obtained from
many different sources like web documents and news articles tend to grow in size as more data becomes available, 
and thus graphs with billions of vertices and edges have become prevalent in practice.
Running the graph algorithms on the large and complex graphs in an efficient and scalable way 
has become an increasingly important and yet challenging problem.
This raises an imperative  need to identify computer architectures that are best suited for solving
graph theoretic problems for large complex graphs.

There are many aspects to consider in an architectural evaluation.
An ideal architecture should provide high performance and good scalability for target applications.
It should also be energy-efficient, cost-effective, and easy to program.
In addition, it should offer reliable, large-capacity storage subsystems
in order to process large data sets, for graph theoretic applications in particular.
These aspects must be carefully weighed in when evaluating various machine architectures,
in conjunction with the run-time behavior and resource usage pattern of the target applications.
Benchmarks have been the most commonly used architectural evaluation tool, since they are
specifically designed to represent the key characteristics of the target algorithms
and hence correctly reproduce their run-time behavior, 
A well-designed benchmark is critical to the accurate architectural evaluation.

Although innumerable benchmarks have been developed for a wide spectrum of applications~\cite{Bailey93,tpc,linpack},
little attempts have been made in developing benchmarks for graph-theoretic applications.
Many existing graph-theoretic benchmarks consist of graph data sets designed mainly for the algorithmic evaluation of  
specific graph algorithms~\cite{johnson1996:cliques,BHOSLIB}.
Recently, a new benchmark suite, called DARPA High Productivity Computer Systems (HPCS) 
Scalable Synthetic Compact Application (SSCA) Graph Analysis Benchmark
(commonly known as SSCA\#2 benchmark), was developed~\cite{bader2005:design}.
The SSCA\#2 benchmark suite is comprised of a synthetic scale-free graph generator~\cite{Chakrabarti04r-mat:a} 
and four kernels each of which
is designed based on a small set of fundamental graph-related algorithms.
The benchmark has found some successes as it is the only benchmark currently available 
designed specifically for architectural evaluation for graph-theoretic problems.
However, the SSCA\#2 benchmark has some notable drawbacks that make the benchmark inadequate to use in any rigorous performance study.
First, the benchmark is not a complete representative of  a wide variety of commonly used graph algorithms and therefore fails to
provide comprehensive models for their algorithmic behavior. 
Some of the kernels in the SSCA\#2 benchmark are not graph-related, but closer to to 
graph construction and min-max finding.
Remaining kernels, although they certainly address fundamental and very important graph problems, covers
only a fraction of existing graph algorithms.
Second, the design flaws in some of the kernels prevent the benchmark from  accurately modeling the real execution-time
characteristics of the targeted algorithms. For example, 
its kernel calculating the betweenness centrality 
scores~\cite{brandes01:afaster} approximates the calculation by finding 
shortest paths between only randomly selected vertices due to high computation cost.
This may result in inaccurate betweenness measures, and more importantly, this may alter the real run-time memory access characteristics of
the original algorithm, since the subgraph formed by the shortest paths found may have significantly different structures 
from the original graph.


We address these issues and propose a new graph-theoretic benchmark.
The benchmark is comprised of a graph generator and a suite of kernels. The graph generator 
synthesizes scale-free graphs using a very fast algorithm based on preferential 
attachment method~\cite{Chakrabarti04r-mat:a,yoo2006:parallel}.
The kernels are comprehensive and designed to represent  a wide range of important graph algorithms, 
including traditional graph algorithms such as search, combinatorial optimization, 
and metrics computation and graph-mining methods, like subgraph clustering (also known as community
finding) and spectral graph algorithms, that have gained more popularity in recent years.
The benchmark is designed to accurately model the run-time characteristics of target algorithms with
minimal computational cost.
Furthermore, its specific and detailed design allows the objective evaluation of the strengths and 
weaknesses of different machine architectures and programming models.




The paper is organized as follows. 
Section~\ref{preliminaries} describes some preliminaries and 
graph representations used in this paper.
The proposed benchmark is described in Section~\ref{benchmark} in greater detail. 
Related work is reported in Section~\ref{relatedwork} followed
by concluding remarks in Sections~\ref{conclusions}.





\section{Preliminaries}
\label{preliminaries}
\subsection{Definitions}
The proposed benchmark uses undirected, weighted, colored graph $G = (V, E)$, 
where $V$ is the set of vertices and $E$ is the set of edges.
Each vertex is given an integer that uniquely identifies the vertex.
The vertices are  colored in gray scale, where the darkness of the vertex is
controlled by a real value (e.g., 0 for {\em white} and 1 for {\em black}).
That is, a vertex $v_i \in V$ is a pair $\langle i, c_i\rangle$, where $0 \le i < |V|$ and
$0 \le c_i \le 1$. Here, $i$ and $c_i$ represent the ID and the color of the vertex $v_i$.
An edge $e = \langle u, v, *\rangle$ is said to be {\em incident} to $u$ and $v$.
The {\em degree} of a vertex $v$ is defined as the number of (undirected) edges that 
are incident to $v$.
An edge $e_i \in E$ is the tuple $\langle u, v, w_i \rangle$, where $u, v \in V$  and
$w_i$ is the weight of the edge such that $0 \le w_i \le 1$. 
The graph is undirected iff for every $\langle u, v, w \rangle \in E$, there exists 
$\langle v, u, w \rangle \in E$.
There are no self-loops in $G$. That is, for every $\langle u, v, w \rangle \in E$, 
$u \neq v$. Also, the $G$ does not have multiple edges between any two vertices $u$ and $v$.
A graph $G' = (V', E')$ is a subgraph of $G$, when $V' \subseteq V$ and $E' \subseteq E$.
A subgraph $G' = (V', E')$ is called a {\em clique}, if for any two vertices $u, v \in V'$,
$\langle u, v, *\rangle \in E'$.

\subsection{Graph Representations}
As stated earlier, the performance of a typical graph algorithm is mainly governed  by
its memory access pattern. Therefore, the graph representation has major impact on
the performance of the graph algorithms. 
Since there can be a large number of graph representations and each of the representations 
has different memory behavior, it is very important to utilize
the most appropriate graph representation schemes to accurately model the run-time behavior of 
the graph algorithms.

We adopted two of the most common graph representations,  {\em adjacency list} and {\em compact
sparse row} (CSR), to store graphs in our benchmark.
We chose these graph representations methods for several reasons.
First, because it is critical for our benchmark to be used to evaluate a variety of machine architecture
with different memory capacities, we have excluded those representations, such as {\em adjacency matrix},
that usually consume a lot of memory space to store given graph.
Furthermore, the graph of interest in this research, scale-free graph, is a sparse graph, and
hence we mainly focused on the graph representations ideal for sparse graphs.
Second, these are two of the most widely used  methods for graph representation.
Finally, these two methods represent graph representations at the extreme ends of spectrum  of
memory access patterns, where the linked list representation exhibits more random memory accesses than
the CSR representation.
We believe using these two widely different graph representations
will provide users with ability to measure the best- and worst-case 
memory performance of the target architecture.

\begin{figure}[]
\centering
\begin{tabular}{c}
\mbox{\epsfxsize=5in\epsfbox{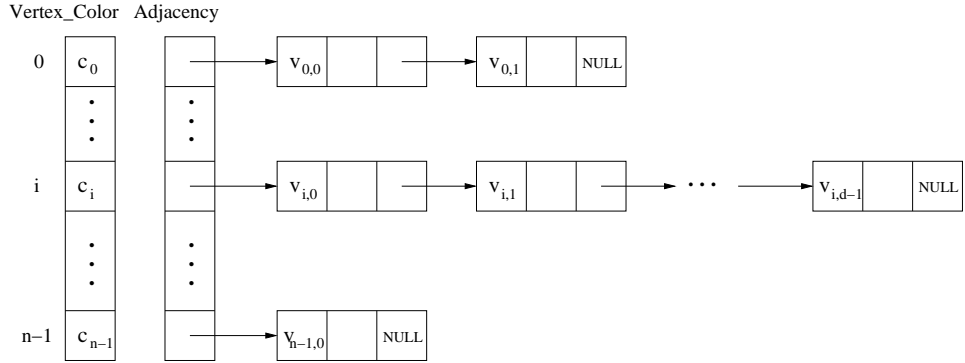}} \\
\end{tabular}
\caption{Adjacency list graph representation for graph $G = (V, E)$, where $n$ = $|V|$.}
\label{linked_list}
\end{figure}

The adjacency-list graph representation is shown in Figure~\ref{linked_list}.
As Figure~\ref{linked_list} shows, the color and adjacent vertices of the vertices are stored in two 
arrays,  {\tt Vertex\_Color} and {\tt Adjacency}. 
The color of a vertex $v_i$ is stored in {\tt Vertex\_Color[i]}. 
The adjacent edges and vertices of $v_i$ are maintained as a linked list.
{\tt Adjacency[i]} points to the first element in the linked list.
Each node in the linked list contains the ID of a vertex, $u$, adjacent to $v_i$ and the weight of edge that
connects $v$ and $u$. 
Given a vertex ID, $j$, the adjacent vertices and the weights of corresponding incident edges
can be obtained by simply following the linked list.

This graph representation is not very efficient in terms of storage space, since
an additional space is required to store a pointer for each edge. 
Further, a graph algorithm that processes a graph in this representation is likely to exhibit
random memory access behavior and hence suffers poor performance, since accessing adjacent vertices
needs chasing pointers to objects that are highly likely dispersed in the heap area.
It should be noted, however, that it is never our intention to design the most efficient graph benchmark.
Rather, our goal is to provide a good architectural evaluation tool to users.

\begin{figure}[]
\centering
\begin{tabular}{c}
\mbox{\epsfxsize=5in\epsfbox{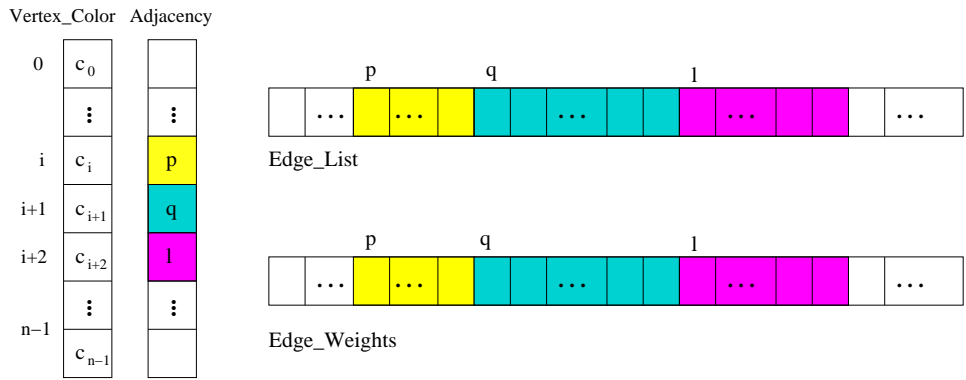}} \\
\end{tabular}
\caption{Compact sparse row (CSR) graph representation for graph $G = (V, E)$, where $n$ = $|V|$.}
\label{CSR}
\end{figure}

The CSR graph representation is one of the most popular graph representation techniques, especially 
ideal for representing sparse graphs such as scale-free graphs.
Its popularity can be attributed to the fact that it requires minimal storage space to store 
graphs while reducing random  memory accesses.
The CSR graph representation is depicted in Figure~\ref{CSR}.
In the CSR graph representation, all the edges and their associated weights are stored in two arrays,
{\tt Edge\_List} and {\tt Edge\_Weights}. Therefore, the size of these arrays is equal to the total number of 
edges in the graph (i.e., $|E|$).
The adjacency information is maintained through another array, {\tt Adjacency}.
Unlike the adjacency list representation, where each element in the {\tt Adjacency} array
points to list of edges that are incident to the corresponding vertex, 
in the CSR representation an entry in the {\tt Adjacency} array  maintains the starting positions
in the {\tt Edge\_List} array for all the edges that are incident to the corresponding vertex.
In Figure~\ref{CSR}, for example, the adjacent vertices of $v_i$ is stored in
{\tt Edge\_List[$p$]} $\ldots$ {\tt Edge\_List[$q-1$]}, and 
the adjacent vertices of $v_j$ is stored in {\tt Edge\_List[$q$]} $\ldots$ {\tt Edge\_List[$l-1$]}, and so on.
Also, the degree of the vertex $v_i$ is $q-p$.
The vertex colors are also stored in {\tt Vertex\_Color} as in the same was as in the adjacency list 
representation.

\begin{figure}[]
\centering
\begin{tabular}{cc}
\mbox{\epsfxsize=2.75in\epsfbox{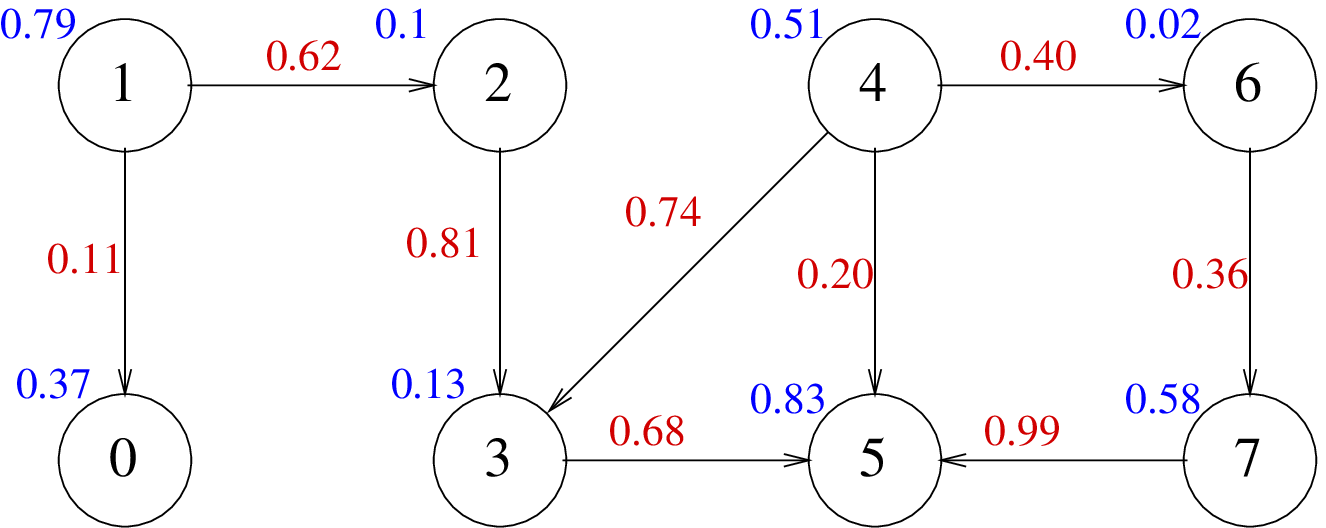}} &
\mbox{\epsfxsize=2.75in\epsfbox{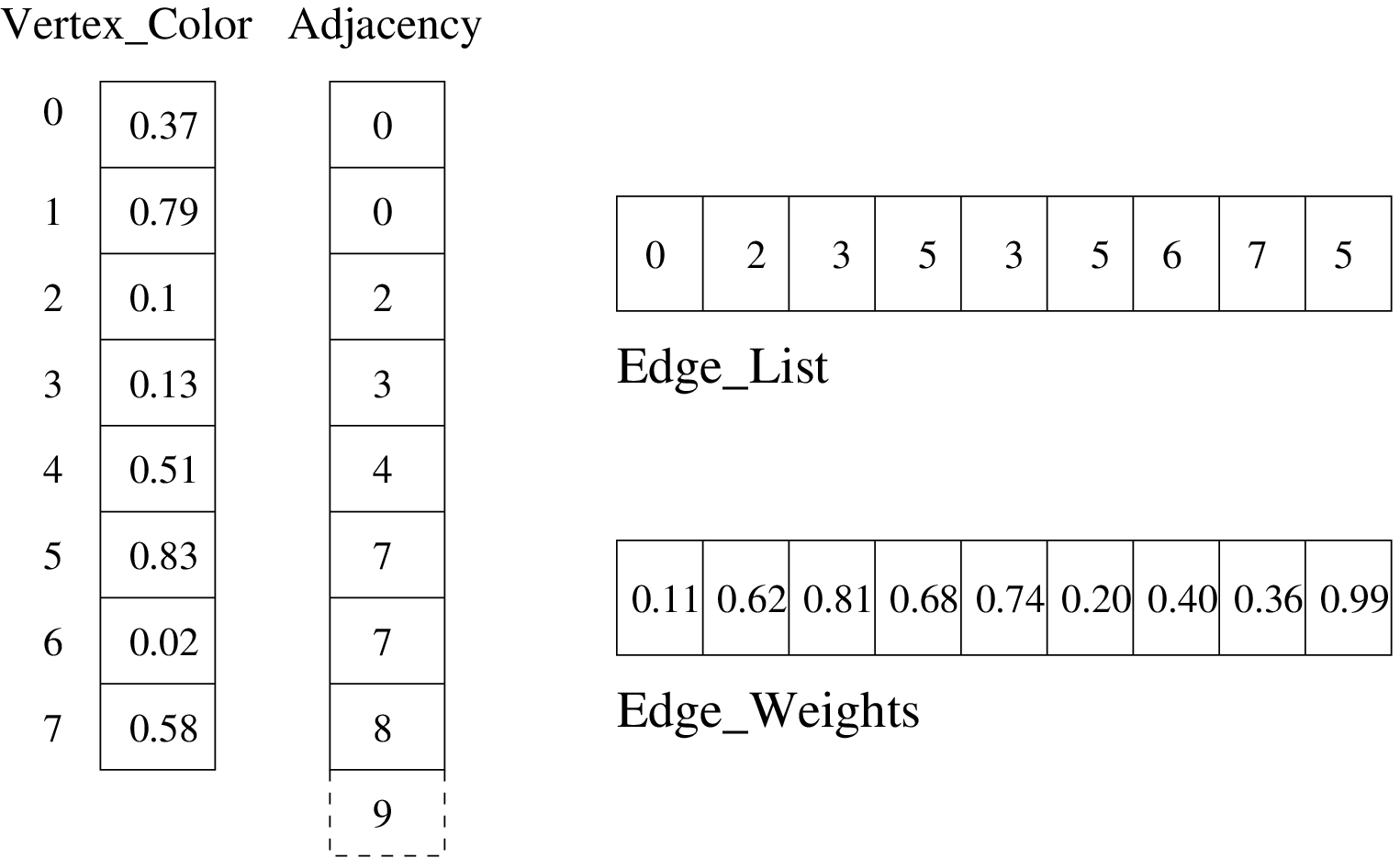}} \\
(a) Sample directed graph &
(b) The corresponding CSR representation \\
\end{tabular}
\caption{A sample graph and its corresponding CSR representation.}
\label{CSR_example}
\end{figure}

Figure~\ref{CSR_example} presents a sample graph and its corresponding CSR graph 
representation\footnote{The adjacency list representation is not shown here, because it is
relatively straightforward to represent the graph in this data structure.
Interested readers should refer to \cite{cormen2001:introduction}}.
The vertex colors and edge weights are depicted in blue and red in Figure~\ref{CSR_example}.a, 
respectively.
A directed graph is used in the example for the simplicity of presentation.
In an undirected graph, each (undirected) edge is translated into two edges in reverse direction
and represented in the same way as the directed graph case.
In this example, vertices 0 and 5 do not have any outgoing edges. This is indicated by 
the fact that the entries in the {\tt Adjacency} array for these vertices have the same value as
the ones immediately following them.
It should be also noted that the {\tt Adjacency} array keeps one additional entry at its end
basically to indicate the array boundary for the last vertex (vertex 7 in the example).

\section{Graph Generator}
\label{generator}
Recent studies have revealed that many important real-world graphs belong to a special class
of graphs called {\em scale-free graphs}. Some of these scale-free graphs
include World-Wide Web~\cite{broder:2000}, 
Internet~\cite{faloutsos99powerlaw,govindan00:heuristics,satorras2001:dynamical,vazquez2002:large}, 
electric power grids~\cite{watts98:collective},
citation networks~\cite{krapivsky2000:connectivity,redner1998:how,tsallis2000:are,vazquez2001:statistics}, 
telephone call graphs~\cite{aiello2000:arandom}, and e-mail network~\cite{ebel2002:scale-free}.
It is very important to understand the behaviors of different graph algorithms on the 
key real-world graphs, and hence, the proposed benchmark includes a graph generator that constructs
synthetic graphs that possess key characteristics of the real-world scale-free graphs~\cite{yoo2006:parallel}.

The graph generator is based on well-known preferential attachment graph model~\cite{barabasi1999:emergence} and
offers significant advantages over the R-MAT based graph generator~\cite{Chakrabarti04r-mat:a} of the SSCA\#2 benchmark. 
Our graph generator has low computational complexity of $O(m)$, compared to
$O(m\cdot log m\cdot log n)$  of the graph generator in SSCA\#2 benchmark, 
where $n$ and $m$ represent the number of vertices and edges of a graph, respectively.
Furthermore, our graph generator was shown to generate more realistic scale-free graphs and can be easily parallelized,
compared to the SSCA\#2 graph generator~\cite{yoo2006:parallel}. 
The preferential attachment method~\cite{dorogovtsev2004}, which our graph generator is based on,
follows well-known generative Barabasi-Albert (BA) model~\cite{barabasi1999:emergence}.
In the BA model, a scale-free graph is constructed by repeatedly creating a
new vertex and then attach it to one of the existing vertices. Here, the existing vertex is selected
with a probability distribution which is proportional to its current degree\footnote{The BA model is also called
rich get richer model, since in this model those rich vertices (vertices with high degree) 
have higher probability to be connected to newly created vertices and hence become richer.}.
The graph generation algorithm is described in Algorithm~\ref{alg:1} in detail.

\begin{algorithm}[plain]
\caption{Graph Generation Algorithm}
\label{alg:1}
\begin{algorithmic}[1]

\STATE // Create a graph G(V, E), where V and E are sets of vertices and edges of G
\STATE // $L$ is list of edges

\STATE Create a clique $C$ of size $c$
\FORALL {$\forall$ edges $(u, v) \in C$}
\STATE  $L \leftarrow L \cup \{u, v\}$
\ENDFOR

\FOR   {$i \leftarrow c$ to $|V|-1$, where $|V|$ denotes the number of vertices in $G$ }
\STATE    id[$u$] $\leftarrow i$ 
\STATE    color[$u$] $\leftarrow$ uniform(0, 1)
\STATE    $d \leftarrow $ uniform(1, 2$\cdot$D),  where $D$ is average vertex degree of $G$
\FOR      {$j \leftarrow 0$ to $d-1$}
\STATE		$k \leftarrow $ uniform(0, $|L|$)
\STATE		$v \leftarrow L_k$
	\STATE         	weight[($u, v$)] $\leftarrow$ uniform(0, 1)
	\STATE  	$L \leftarrow L \cup \{u, v\}$
	\ENDFOR
	\ENDFOR

	\end{algorithmic}
	\end{algorithm}


The algorithm begins with constructing  a clique of a given size. Vertices in the initial clique will
form a set of high-degree vertices called {\em hubs}. 
Selecting an existing vertex based on its current degree can incur very high computation overhead,
especially for larger graphs, since entire vertex set needs to be scanned for each newly added vertex.
Our graph generator maintains a list that contains the end points of all the edges in partially constructed
graph (steps 5 and 15 in \ref{alg:1}) in order to speed up this selection process. 
The basic idea here is to select a vertex from the list of end points.
Since the number of occurrences of each vertex in the list is equal to its degree, 
the probability that a vertex is selected is proportional to its degree.
This enables the creation of edges in a constant time and reduces the overhead considerably.

The degree of a newly added vertex follows a uniform distribution with given average vertex degree 
as its mean (step 10). This is to ensure that the generated graph does not have uniform structure.
Each edge in the graph is associated with a weight between 0 and 1, and similarly, each vertex in the graph
has a color, which is also a real value between 0 and 1 (steps 9 and 14). 
These values are used by some of the kernels in the benchmark, as explained later.

Parallelizing the graph generator is also very simple.
In the parallel graph generator, each processing element (PE) (e.g., processes or threads) maintain
its own list of end points in the same manner as in the sequential generator. The source 
vertices of the edges in a list are owned by a processing element that maintains the list.
However, the destination vertices may belong to other processing elements. When the generation local edges
is complete, a final graph synthesize by merging all the edges with the same destination end points.
Since processing elements create local edges independent from each other, this algorithm is
embarrassingly parallel.
Interested readers should refer~\cite{yoo2006:parallel} for more detailed description of the parallel
graph generator.

\section{Description of Graph Benchmark}
\label{benchmark}
We designed the proposed benchmark with a set of clear design goals.
First, the benchmark should be comprehensive in that it should cover a set of fundamental and
more commonly used graph algorithms so that it can present  a collection of varying
resource usage and execution patterns typical graph algorithms exhibit during their execution.
Second, the benchmark should be designed in such a way that it captures and models
only the essential algorithm characteristic of the target algorithms.
This requires a thorough understanding of target algorithms.
Third, the benchmark should be very specific about algorithms, graph representations, and all data structures
used by the algorithms so that ones can make unbiased evaluation of target architecture.
Finally, the benchmark should be computationally tractable so that runs can finish in a
reasonable amount of time.

As the first step in designing the proposed  benchmark, we surveyed the
graph algorithms that are commonly used in various important applications in practice.
The survey was performed to make out benchmark as comprehensive as possible by ensuring
that the new benchmark covers an extensive list of key graph algorithms.
Furthermore, this study should help us to great extent better understand and correctly capture
the run-time behavior of the target graph algorithms.
The surveyed graph algorithms are classified into 5 groups\footnote{They certainly
can be classified into more groups.  The classification is done to meet our design goals,
especially the computational and space constraints. These have excluded some of the graph algorithms.}. 
The algorithms in each category are modeled, mainly focusing on their
algorithmic and run-time characteristics, and translated into a kernel.
These kernels of the proposed benchmark are described in detail in the following.
The detailed description of the proposed benchmark design is given below.

\subsection{Kernel 1: Graph Search}
\subsubsection{Graph search algorithms}
Algorithms in this class pertain to traversing paths in given graph and finding the
information associated with the paths and (typically) vertices on the paths.
These algorithms are fundamental graph algorithms that have been
applied to solving a variety of scientific and engineering problems.
In this research, we surveyed many classical graph algorithms that perform a form of
graph search, including  breadth-first search (BFS), depth-first search (DFS),
Dijkstra's  algorithm,  Prim's algorithm, Kruskal's algorithm, and Bellman-Ford algorithm.

We will not discuss these algorithms in further detail in this paper, 
as the detailed description of these algorithms
is widely available in the literature~\cite{cormen2001:introduction},
but we should point out  key characteristics of these algorithms governing their run-time behavior.
Although they deal with the vertices and edges in a given graph, the primarily operate on {\em sets}
obtained from the graph. The BFS algorithm, for example, accesses two sets, which maintain the
vertices at the current level and all the visited vertices, when it expands the search to 
the next level.
Similarly, other algorithms in this class also access elements in sets in one way or
another.
Here, the way the elements in the sets are maintained and accessed dictates
the run-time memory access patterns of these algorithms.
If a set is unordered (as in the case of BFS), then most likely, accesses to the elements in the sets
can be done sequentially, while
accesses to ordered sets involve more random memory accesses.
In addition to the set operations, adjacency list, accesses to which are usually in random order,
also affects their run-time memory behavior.

It has been generally believed that graph algorithms, especially the graph search algorithms,
have poor cache utilization due to their random memory accesses.
Use of efficient data structures to maintain the sets (e.g., {\em priority queues})
and represent graphs (e.g., {\em compressed sparse row} (CSR) graph representation), however,
may allow the search algorithms to access the memory in more sequential manner, in contrast to
the popular belief.
We aim at capturing this behavior correctly in this kernel.

\subsubsection{Benchmark description}
Algorithm~\ref{kernel:1} describes the kernel 1 in detail.
The design of the kernel is  essentially based on the {\em single-source shortest paths} 
problem~\cite{cormen2001:introduction}.
Here, the length of a path is the sum of weights of the edges that comprise the path.
The input to the kernel is graph, $G(V, E)$ to be searched and the source from which search begins.
The kernel returns an array of size $\mid$V$\mid$, where $i$-th entry has the length of path
from $s$ to $v_i$.
There can be many variations of
algorithms to find all the shortest paths from a single source, but
the proposed kernel combines the behavior of the  BFS algorithm
and ordered set accesses.

A queue of vertices is used to maintain a set of unvisited vertices and denoted as $Q$.
Since the algorithm needs to find a vertex with the smallest current path length from $s$ 
(line 10 in Algorithm~\ref{kernel:1}), 
the elements in $Q$ should be maintained in an increasing order of their current path lengths.
That is, $Q$ is an ordered set.

When a vertex $v$ is selected from $Q$, its path length is
the minimum path length from $s$ to $v$ and will not be improved by any other paths.
Hence, the vertex $v$ is removed from $Q$.
Then, all of its neighboring vertices are added to $Q$, if they are not already in $Q$,
followed by relaxing their path lengths.
The {\em relaxation} of a vertex basically updates the current length of path to the vertex,
if there exists a path of shorter length to the vertex~\cite{cormen2001:introduction}. The relaxation is 
performed in lines 26 to 30 in Algorithm~\ref{kernel:1}.

%
\begin{algorithm}[plain]
\caption{Kernel 1: Graph Search}
\label{kernel:1}
\begin{algorithmic}[1]

\STATE Input:   $G(V, E)$ and a vertex $s$
\STATE Output:  Array W, where W[$v$] is the weight of path from $s$ to $v$

\STATE Create an array W of size $\mid$V$\mid$
\FORALL {$\forall u \in V$}
\STATE W[$u$] $\leftarrow \infty$
\ENDFOR

\STATE W[s] $\leftarrow$ 0
\STATE $Q$ $\leftarrow$ \{s\}

\WHILE {$Q$ $\neq$ $\emptyset$}
\STATE $u \leftarrow$ Extract\_Min($Q$, W)
\FORALL{$\forall v \in$ Adjacent($u$)}
\IF {$v \not\in Q$}
\STATE $Q$ $\leftarrow$ $Q$ $\cup$ \{$v$\}
\ENDIF
\STATE $w \leftarrow$ weight of $e(u, v)$
\STATE Relax(W, $u, v, w$)
\ENDFOR
\ENDWHILE

\STATE \vspace{0pt}
\STATE {\bf function} Extract\_Min(Q, W)
\STATE $v \leftarrow v^\prime\in Q$ such that W[$v^\prime$] $\le$ W[$u$] for $\forall u\in Q$
\STATE $Q \leftarrow Q - \{v\}$
\STATE return $v$
\STATE {\bf end function}

\STATE \vspace{0pt}
\STATE {\bf function} Relax(W, u, v, w)
\IF {W[v] $>$ W[v] + Weight(u, v)}
\STATE W[v] $\leftarrow$ W[u] + Weight(u, v)
\ENDIF
\STATE {\bf end function}

\STATE \vspace{0pt}
\STATE {\bf function} Adjacent(u)
\STATE return $\{v\mid (u, v, *) \in E\}$
\STATE {\bf end function}


	\end{algorithmic}
	\end{algorithm}


\subsection{Kernel 2: Spectral Graph Analysis}
\subsubsection{Spectral graph problem}
The graph algorithms in this class of algorithms
are related to the spectral graph theory~\cite{cvetkovic80,cvetkovic88,cvetkovic97}.
Essentially, the spectral graph theory is the study of properties of a graph in relationship to the
eigenvalues and eigenvectors of its corresponding adjacency matrix.
In other words, the algorithms in this class treat a graph as a matrix and aim at finding
the eigenvalues/eigenvectors of the matrix.
Prominent exemplary algorithms in this class include {\em page rank}~\cite{Page99thepagerank} and
{\em random walk with restart} (RWR)~\cite{tong2008:Random}.
These algorithms compute eigenvectors, each value in which is typically interpreted as the relative importance or
relevance of the corresponding vertex to other vertices in the graph.

The eigenvalue/eigenvector problem has been studied quite extensively and a number of eigensolvers
are available~\cite{trilinos,hernandez2005:SLEPc}. Also, it is well known that the algorithms in this class
generally do not scale well and their performance and scalability
are greatly affected by graph representation method.
We attempt to model the essence of the run-time characteristics of these  algorithms in this kernel.

\subsubsection{Benchmark description}
The proposed kernel relies on the simple {\em power method}~\cite{ipsen2005:analysis} to find
an eigenvalue for given graph.
The power method finds the largest (dominant)  eigenvalue and its corresponding eigenvector.
Though simple, the power methods works pretty well for well-conditioned sparse graphs
such as the graphs targeted by the proposed benchmark.
This decision is justified by our design objective to model the behavior of the algorithms in the domain of the
spectral graph theory while guaranteeing the completion of the computation in a reasonable time.

The input to the kernel are graph $G(V, E)$ and threshold values that basically used to
limit the number of iterations. The kernel produces an eigenvector for the largest eigenvalue as output.
The eigenvector is initialized with the same constant, but any random numbers can be used.
The function Multiplication in lines 22 to 24 in the Algorithm~\ref{kernel:2} 
performs a matrix-vector multiplication.
The convergence is determined by the condition, $\frac{\parallel X^{i+1}-X^{i}\parallel}{\parallel X^{i}\parallel}$.
Here, $\parallel X\parallel$ denotes the $L_1$ norm of $X$.
If the ratio is smaller than given threshold, $\epsilon$, then the iteration stops.
Because the power method can converge very slowly, we limit the number of iterations by another
input parameter, $m$, which represents the maximum number of iterations allowed.

It should be noted that  implementation of this kernel must not convert given graph representation 
into an equivalent {\em adjacent matrix} representation. The reason being for this restriction is that
although the adjacency matrix graph representation will facilitate  certain operations in the
kernel (i.e., lines 16 to 17 in the Algorithm~\ref{kernel:2}), accessing the matrix will have
memory access patterns that are very different from the accesses to other graph representations.

\begin{algorithm}[plain]
\caption{Kernel 2: Spectral Graph Analysis}
\label{kernel:2}
\begin{algorithmic}[1]

\STATE Input:   $G(V, E)$, $\epsilon$ and $m$
\STATE Output:  An array $X$

\STATE Let $n$ = $\mid$V$\mid$
\STATE Create an array C of size $n$
\FORALL {$\forall i$, where $i$ = 0 $\ldots$ $n-1$}
\STATE C[$i$] = $k$, such that $k$ = $\sum_{j=0}^{n-1}$  Weight($i$, $j$)  
\ENDFOR
\FORALL {$\forall j$, where $j$ = 0 $\ldots$ $n-1$}
\STATE $X_{j}^{0} \leftarrow \frac{1}{n}$
\ENDFOR

\STATE $t \leftarrow$ 0
\STATE $i \leftarrow$ 0
\STATE $\theta \leftarrow \infty$
\WHILE{$t < m$ and $\theta > \epsilon$}
\STATE $X^{i+1} \leftarrow$ Multiplication($X^i$)
\STATE $X^{i+1} \leftarrow \frac{X^{i+1}}{\parallel X^{i+1}\parallel}$
\STATE $\theta \leftarrow \frac{\parallel X^{i+1} - X^{i}\parallel}{\parallel X^{i}\parallel}$

\STATE $i \leftarrow i+1$
\STATE $t \leftarrow t+1$
\ENDWHILE

\STATE \vspace{0pt}
\STATE {\bf function} Multiplication($X$)
\STATE return $A$ such that $A_k$ = $z$, where $z$ = $\sum_{i=0}^{n-1}\sum_{j=0}^{n-1} w\cdot X_j$ and ($i, j, w$) $\in$ E 
\STATE {\bf end function}

	\end{algorithmic}
	\end{algorithm}


\subsection{Kernel 3: Vertex and Edge Accesses}
\subsubsection{Adjacency finding}
The most frequently  used graph operation is without doubt, given a
vertex,  finding a set of its adjacent vertices.
In fact, most of the graph algorithms need to access adjacency data.
The graph search algorithms, for example,
must access adjacent vertices to a given vertex to traverse paths.
Those algorithms that explore the graph structure, such as maximal clique finding and
subgraph pattern matching, also need the vertex adjacency.
These algorithms typically access the data structures that store the adjacency information repeatedly
to find neighboring vertices.
Therefore, accesses to the adjacency information dominates the run-time memory access
patterns of many graph algorithms, especially  the graph algorithms that operate primarily on the adjacency data.
The repeated adjacency finding operations usually result in a memory access behavior that is characterized by
a combination of random and sequential memory accesses.

A graph transformation algorithm called {\em graph hierarchicalization}~\cite{abello2002:MGV} is
of particular interest because this algorithm repeatedly accesses the adjacency information,
and hence, our kernel for the vertex and edge accesses is modeled based on this algorithm.
The graph hierarchicalization basically transforms a graph into a hierarchy of
graphs, where the graph at each level provides a view of the graph in different granularity.
More specifically, the graph hierarchicalization algorithm
gradually converts the given graph into a graph of smaller size by
iteratively coalescing vertices and edges of the given graph.
Mapping information is also created to  map the vertices and edges in two adjacent graphs in the
hierarchy, although this is not included in this kernel.  
Since the graph hierarchicalization can abstract and reduce the size of a graph,
it is mainly used for graph summarization and visualization.

Another aspect we try to capture by this kernel is the  behavior of the algorithms that
require frequent updates to the given graph. 
We incorporated the updates by replacing the coalesced vertices that with a (mega) vertex that 
represent the coalesced vertices and reconnecting appropriate edges in this kernel.

\subsubsection{Benchmark description}
Input to the kernel is  graph $G(V, E)$ and a constant $\gamma$, which is used to
control the number of coalescing performed.
Basically, the $\gamma$ is used to compute the number of vertices from which
the coalescences start.
As stated above, the number of coalescences is determined by $\gamma$, which in turn is
used to compute the number of vertices with which the coalescences are performed (lines 3 and 7).

In each coalescence, a vertex selected from the graph in random.
Then, the vertex is coalesced with its adjacent vertices. The coalesced vertices are
kept in $S$. 
Also, the average color of the vertices in $S$ is computed for the new vertex that represents
$S$. 
The lines from 13 to 20 rearranges the edges that are connected to those vertices in $S$ in such a way
that they are connected to the new vertex for $S$. Once coalescing vertices and reconnecting
edges, all the vertices in $S$ are replaced by the new vertex representing $S$ (lines from 21 to 24).
Since this kernel alters the structure of the original graph, it is necessary to keep track of
the changes made so that the original graph can be restored (line 27). It is allowed to save the 
entire graph disk, if bookkeeping of the changes consume too much resources in terms of time and space. 

\begin{algorithm}[plain]
\caption{Kernel 3: Vertex and Edge Accesses}
\label{kernel:3}
\begin{algorithmic}[1]

\STATE Input:   $G(V, E)$ and $\gamma$ (reduction ratio)
\STATE Output:  G$^\prime$(V$^\prime$, E$^\prime$), where G$^\prime$ is a new coalesced graph 

\STATE $n\leftarrow\gamma\cdot\mid V\mid$
\STATE $i \leftarrow$ 0
\STATE V$^\prime$ $\leftarrow$ V
\STATE E$^\prime$ $\leftarrow$ E
\WHILE {$i < n$}
\STATE Let $u$ be vertex randomly selected from V$^\prime$
\STATE $S \leftarrow \{u\}$ + Adjacent($u$)
\STATE Let $u^\prime$ be a new vertex representing $S$
\STATE $c \leftarrow \sum_{v}^{}$Color[$v\in S$]/$\mid S\mid$ 
\STATE Color[$u^\prime$] $\leftarrow c$
\FORALL {$\forall v \in S$}
\FORALL {$\forall e = (v_1, v_2, w) \in$ E$^\prime$ such that $v = v_1$ }
\STATE  Update $e = (v, v_2, w)$ 
\ENDFOR
\FORALL {$\forall e = (v_1, v_2, w) \in$ E$^\prime$ such that or $v = v_2$}
\STATE  Update or $e = (v_1, v, w)$
\ENDFOR
\ENDFOR
\FORALL {$\forall v \in S$}
\STATE E$^\prime$ $\leftarrow$ E$^\prime$ - $\{v\}$ 
\ENDFOR
\STATE E$^\prime$ $\leftarrow$ E$^\prime$ $\cup u$$^\prime$
\STATE $i \leftarrow i + 1$
\ENDWHILE
\STATE Restore G
	\end{algorithmic}
	\end{algorithm}


\subsection{Kernel 4: Graph Metric}
\subsubsection{Graph metric computation}
Various metrics are measured and utilized in the area of graph mining and graph structure analysis.
Some of common graph metrics include degree distribution, clustering coefficients, and modularity.
Clustering coefficient of a vertex, 
proposed by Watts and Strogatz~\cite{Watts98}, is a strong indicator of how densely 
the vertex is connected with
its neighboring vertices and can be used to identify the centroids of potential clusters~\cite{clauset-2005-72}.
Similarly, the modularity~\cite{newman:2004a} is a metric that directly indicates the density of
a subgraph and is widely used in many community finding 
algorithms~\cite{newman:2001b,newman:2001c,newman:2004a,newman:2004b,newman:2004c,newman:2004d,girvan-2002-99:community}.
The betweenness centrality~\cite{brandes01:afaster} of a vertex, on the other hand, is a good indication of
the likelihood of the vertex belonging to a cluster.

Computation of these graph metrics generally shows random memory access patterns, since
it is closely related to and heavily relies on adjacency finding.
We selected one of the most popular ans simplest graph metrics, clustering coefficients, as a base model
in designing this kernel.

\subsubsection{Benchmark description}
This kernel takes a graph $G(V, E)$ and a constant $m$ as its input.
The constant $m$ is used to control the number of vertices for which the kernel finds
clustering coefficients.
The computation of the clustering coefficient (in lines 6 to 16 in Algorithm~\ref{kernel:4} 
is straightforward and we will omit its  description.
It should be noted, however, that the kernel counts each edge as a directed edge in its
calculation. This is because each undirected edge is represented as two directed edges of
reverse direction in the graph representations used in our benchmark.

\begin{algorithm}[plain]
\caption{Kernel 4: Graph Metric}
\label{kernel:4}
\begin{algorithmic}[1]

\STATE Input:   $G(V, E)$ and $m$ 
\STATE Output:  An array $CC$ that contains the clustering coefficients of $m$ vertices

\STATE Create array $CC$ of size $m$
\STATE $i \leftarrow$ 0
\FORALL{$i < m$}
\STATE Let $u \in $V be a vertex randomly selected from G
\STATE $S \leftarrow $ Adjacent($u$)
\STATE $c \leftarrow$ 0
\FORALL {$\forall v \in S$}
\STATE $S^\prime \leftarrow$ Adjacent($v$)
\FORALL {$\forall v^\prime \in S^\prime$}
\IF {$v^\prime \in S$}
\STATE $c \leftarrow c$ + 1
\ENDIF
\ENDFOR

\STATE $CC$[$u$] $\leftarrow\frac{c}{\mid S\mid\cdot(\mid S \mid - 1)}$
\ENDFOR
\STATE $i \leftarrow i$ + 1
\ENDFOR

	\end{algorithmic}
	\end{algorithm}


\subsection{Kernel 5: Global Optimization}
\subsubsection{Global optimization problem for graphs}

There are a class of graph algorithms that finds solutions by optimizing
certain objective functions.
Many NP-Complete graph algorithms that solve combinatorial optimization problems belong to this class of
algorithms, although they are not of interest in this research mainly due to their
high computational complexity.
We are mainly interested in those graph algorithms which follow greedy method~\cite{cormen2001:introduction},
These algorithms compute solutions by iteratively constructing a sequence of (partial) solutions.
At each iteration, a new solution is constructed in a way to improve given objective function.
A well-known graph partitioning algorithm MeTis~\cite{karypis95metis}, for instance,
partitions a given graph into a fixed
number of partitions in such a way to minimize the cuts between partitions to reduce the communications
on message-passing parallel computers.
A classic community finding algorithm by Newman and Girvan
constructs a dendrogram of vertices that results in the largest modularity value~\cite{newman:2004a}.

The objective functions employed by these algorithms are usually global measures that require
access and process entire graph to compute. This global accesses in fact make
it very difficult to parallelize this class of algorithms.
The algorithms in this class are modeled in our benchmark to capture this algorithmic behavior.
Our benchmark models a well-known community finding algorithm called AUTOPART~\cite{chakrabarti2004:autopart}.
This algorithm iteratively divides a given graph into a set of communities
in such a way that the value of an entropy-based objective function decreases.
The AUTOPART algorithm is an ideal representative algorithm for this class of algorithms to model,
because it incurs relatively low computational overhead while exhibiting the run-time behavior common to
the algorithms in this class.
%
%
%

\subsubsection{Benchmark description}
The input to the kernel is graph $G(V, E)$ and two constants, $\alpha$, and $m$.
The $\alpha$ is mainly used to reduce the search space by filtering out a set of vertices
whose color is less than $\alpha$ (line 3 in Algorithm~\ref{kernel:5}).
The $m$ is used to control the number of iterations to ensure the computation finishes within
a reasonable among of time.

The kernel basically partitions the vertices of $G$ into a set of clusters. 
The vertices are grouped together in such a way that the grouping will improve the
overall {\em density} of groups.
The density of a group is defined as the ration of the number of internal edges to that of 
external edges. Here an internal edge refers to edge that connects two vertices in the given group of
vertices. External edge, on the other hand, is an edge one of whose endpoints does not belong to the
group.

The kernel maintains a set, $C$, of groups of vertices. 
It first selects a group $g \in C$ that has the minimum density value. Then, it tries to split $g$ into 
two by iteratively checking moving a vertex from $g$ to the new group will improve the
overall density (lines from 12 to 24).
The objective function used by kernel is defined by a function called {\em Objective} in
the Algorithm~\ref{kernel:5}.
The objective function sums up the value of Shannon's binary entropy function~\cite{shannon51:prediction} for 
each group in $C$. 
The kernel continues the split of $C$ until it cannot split any of the groups in $C$ (lines 18 to 20)
or its iteration reaches given $m$.

\begin{algorithm}[plain]
\caption{Kernel 5: Global Optimization}
\label{kernel:5}
\begin{algorithmic}[1]

\STATE Input:   $G(V, E)$, $\alpha$, and $m$ 
\STATE Output:  $k$, the number of clusters that maximizes objective function

\STATE Let G$^\prime$(V$^\prime$, E$^\prime$) be a graph, where
V$^\prime$ = $\{v\in$ V$\mid$ Color[$v$] $>\alpha\}$ and
E$^\prime$ = $\{e = (u, v, w)\in$E $\mid u, v \in$ V$\}$
\STATE $i \leftarrow 0$
\STATE $k \leftarrow 1$
\STATE $g_1 \leftarrow V^\prime$
\STATE $C_1 = \{g_1\}$
\WHILE{$i < m$}
\STATE $g_{min} \leftarrow g_j$ such that $g_j =  min_{g\in C_i}$ Density($g$)
\STATE $C_i \leftarrow C_i - g_{min}$
\STATE $g_{new} \leftarrow \emptyset$
\FORALL {$\forall v\in g_{min}$}
\STATE $g_{new} \leftarrow g_{new} \cup \{v\}$
\IF {Objective($C_i \cup g_{min}$) $>$ Objective($C_i \cup (g_{min} - \{v\}) \cup (g_{new} \cup \{v\})$)}
\STATE $g_{min} \leftarrow g_{min} - \{v\}$
\STATE $g_{new} \leftarrow g_{new} \cup \{v\}$
\ENDIF
\IF {$g_{new} = \emptyset$}
\STATE Stop
\ELSE
\STATE $C_i \leftarrow C_i \cup g_{min} \cup g_{new}$
\STATE $k \leftarrow k + 1$
\ENDIF
\ENDFOR
\STATE $i \leftarrow i + 1$
\ENDWHILE

\STATE \vspace{0pt}
\STATE {\bf function} Density($g$)
\STATE $S_i \leftarrow \{e = (u, v, w)\in E^\prime\mid u\in g$ and $v\in g\}$
\STATE $S_e \leftarrow \{e = (u, v, w)\in E^\prime\mid e\not\in S_i$ and $(u\in g$ or $v\in g)\}$
\STATE $d \leftarrow \frac{\mid S_i\mid}{\mid S_e\mid}$
\STATE {\bf end function}

\STATE \vspace{0pt}
\STATE {\bf function} Objective($C$)
\STATE return $\sum_{g\in G}$ H(Density($g$))
\STATE {\bf end function}

\STATE \vspace{0pt}
\STATE {\bf function} H($p$)
\STATE return $-p\cdot log_2 p - (1-p)\cdot log_2 (1-p)$
\STATE {\bf end function}

	\end{algorithmic}
	\end{algorithm}


%

\section{Related Work}
\label{relatedwork}
Although there are numerous graph algorithms reported in the literature,
little work has been done in graph-centric benchmark development.
The most relevant work in this area is probably the recently developed 
benchmark suite, called DARPA High Productivity Computer Systems (HPCS) Scalable Synthetic Compact Application (SSCA)
Graph Analysis Benchmark, which is commonly known as SSCA\#2 benchmark~\cite{bader2005:design}.
The SSCA\#2 benchmark suite is comprised of a synthetic scale-free graph generator based on RMAT 
method~\cite{Chakrabarti04r-mat:a} 
and four kernels. Each of the kernels is designed based on a small set of fundamental graph-related algorithms.
It is the only benchmark currently available for architectural evaluation for graph-theoretic problems.
However, the SSCA\#2 benchmark suffers significant shortcomings.
One of the biggest drawbacks is that the benchmark is not comprehensive.
Also, among the four kernels in the benchmark, kernels 1 and 2 are not directly graph-related.
Remaining kernels suffer design flaws that prevent the benchmark from  accurately modeling the real execution-time
characteristics of the targeted algorithms. 
In addition, its use of RMAT as a synthetic graph generator results in longer graph generation time and 
graphs that do not resemble common real-world scale-free graphs.

Gokhale et. al.~\cite{gokhale2008:hardware} have evaluated the use of  
existing and emerging computing architectures, storage technologies,
and programming models to solve data-intensive problems in various scientific and engineering disciplines.
One of the data-intensive applications considered in their research was graph algorithms. In particular, they were
interested in the graph algorithms that access very large graphs stored in external storage devices.
In an effort to investigate the use of the new storage devices and identify what is needed for the efficient and scalable
computation of graph applications, they developed a graph-centric benchmark that operates on out-of-core graphs.
This benchmark measures the performance and scalability of the graph ingestion and search.

A library that contains a set of the most commonly used graph algorithms has 
been developed~\cite{lumsdaine07:_challenges_in_parallel_graph}.
The library, called Boost Graph Library (BGL), enables the reuse of graph algorithms and data 
structures by providing the users with a generic interface that allows access to a graph's structure while
hiding the details of the implementation. 
Therefore, any graph algorithms or libraries that implement this interface will be able to
access the BGL generic algorithms and other algorithms that use the interface.
The BGL offers  some of the fundamental graph algorithms that include
breadth- and  depth-first search algorithms, Dijkstra's shortest path algorithm, Kruskal's and Prim's 
minimum spanning tree algorithms, and find strongly connected components algorithms.
In addition to the graph algorithms, the BGL supports common graph representations: adjacency\_list and
adjacency\_matrix.
The BGL itself can be used as a benchmark, since it supports a wide range of fundamental 
graph algorithms and common data structures to represent graphs.
However, it lacks a means to fine control the behavior of these graph algorithms, which any real
benchmark is expected to provide.

\section{Conclusions}
\label{conclusions}
Graph  has become an increasingly important and widely-used tool in a wide range of
emerging disciplines such as web mining, computational biology, social network analysis, and text analysis.
These graphs are usually very large and have very complex structures.
Running the graph algorithms on these large and complex graphs in an efficient and scalable way 
has become an increasingly important and yet challenging problem, which
raises an imperative  need to identify computer architectures best suited for running
graph algorithms.
There exists, however, no good benchmark that is designed specifically for graph algorithms.

We develop and present a new graph-theoretic benchmark in this paper to address this issue.
The benchmark is comprised of a very efficient graph generator and six kernels. 
We thoroughly studied a large number of traditional and contemporary graph algorithms
reported in the literature to have clear understanding of  their algorithmic and run-time 
characteristics. The developed kernels are comprehensive and correctly model the common behavior of
a wide range of important graph algorithms.
We expect that the developed benchmark will serve as a much needed tool for evaluating different 
architectures and programming models to run graph algorithms.

\bibliographystyle{abbrv}
\bibliography{../references}

\begin{thebibliography}{10}

\bibitem{abello2002:MGV}
J.~Abello and J.~Korn.
\newblock Mgv: a system for visualizing massive multidigraphs.
\newblock {\em Transactions on Visualization and Computer Graphics},
  8(1):21--38, 2002.

\bibitem{aiello2000:arandom}
W.~Aiello, F.~Chung, and L.~Lu.
\newblock A random graph model for massive graphs.
\newblock In {\em Proceedings of the thirty-second annual ACM symposium on
  Theory of computing (STOC '00)}, pages 171--180, New York, NY, USA, 2000.

\bibitem{bader2005:design}
D.~Bader and K.~Madduri.
\newblock Design and implementation of the hpcs graph analysis benchmark on
  symmetric multiprocessors.
\newblock In {\em Proc. 12th Int’l Conf. on High Performance Computing},
  2005.

\bibitem{barabasi1999:emergence}
A.-L. Barabasi and R.~Albert.
\newblock Emergence of scaling in random networks.
\newblock {\em Science}, 286:509, 1999.

\bibitem{linpack}
L.~Benchmarks.
\newblock {\tt http://www.netlib.org/linpack/}.

\bibitem{BHOSLIB}
BHOSLIB.
\newblock Bhoslib: Benchmarks with hidden optimum solutions for graph problems,
  2009.
\newblock {\tt
  http://www.nlsde.buaa.edu.cn/kexu/benchmarks/graphbenchmarks.htm}.

\bibitem{brandes01:afaster}
U.~Brandes.
\newblock A faster algorithm for betweenness centrality.
\newblock {\em Journal of Mathematical Sociology}, 25:163--177, 2001.

\bibitem{broder:2000}
A.~Broder, R.~Kumar, F.~Maghoul, P.~Raghavan, S.~Rajagopalan, R.~Stata, and
  A.~Tomkins.
\newblock Graph structure in the web: Experiments and models.
\newblock In {\em 9th World Wide Web Conference}, 2000.

\bibitem{chakrabarti2004:autopart}
D.~Chakrabarti.
\newblock Autopart: parameter-free graph partitioning and outlier detection.
\newblock In {\em PKDD '04: Proceedings of the 8th European Conference on
  Principles and Practice of Knowledge Discovery in Databases}, pages 112--124,
  New York, NY, USA, 2004. Springer-Verlag New York, Inc.

\bibitem{Chakrabarti04r-mat:a}
D.~Chakrabarti, Y.~Zhan, and C.~Faloutsos.
\newblock R-mat: A recursive model for graph mining.
\newblock In {\em In SDM}, 2004.

\bibitem{clauset-2005-72}
A.~Clauset.
\newblock Finding local community structure in networks.
\newblock {\em Physical Review E}, 72:026132, 2005.

\bibitem{newman:2004d}
A.~Clauset, M.~E.~J. Newman, and C.~Moore.
\newblock Finding community structure in very large networks.
\newblock {\em Phys. Rev. E}, 70(6):066111, Dec. 2004.

\bibitem{cormen2001:introduction}
T.~H. Cormen, C.~E. Leiserson, R.~L. Rivest, and C.~Stein.
\newblock {\em Introduction to Algorithms}.
\newblock {MIT} Press, Cambridge, {MA}, second edition, 2001.

\bibitem{tpc}
T.~P.~P. Council.
\newblock {\tt http://www.tpc.org/}.

\bibitem{cvetkovic80}
D.~Cvetkovi$\acute{c}$, M.~Doob, and H.~Sachs.
\newblock {\em Spectra of Graphs}.
\newblock 1980.

\bibitem{cvetkovic88}
D.~Cvetkovi$\acute{c}$, M.~Doob, and H.~Sachs.
\newblock {\em Recent Results in the Theory of Graph Spectra (Annals of Disrete
  mathematics series)}.
\newblock North-Holland, 1988.

\bibitem{cvetkovic97}
D.~Cvetkovi$\acute{c}$, P.~Rowlinson, and S.~Simi$\acute{c}$.
\newblock {\em Eigenspaces of Graphs}.
\newblock 1997.

\bibitem{Bailey93}
{D. H. Bailey et al.}
\newblock {T}he \mbox{{N}{A}{S} {P}arallel} {B}enchmarks.
\newblock Technical Report NASA Technical Memorandom 103863, {N}{A}{S}{A}
  {A}mes {R}esearch {C}enter, 1993.

\bibitem{dorogovtsev2004}
S.~N. Dorogovtsev and J.~F.~F. Mendes.
\newblock {\em Evolution of Networks}.
\newblock Oxford University Press, 2004.

\bibitem{ebel2002:scale-free}
H.~Ebel, L.-I. Mielsch, and S.~Bornholdt.
\newblock Scale-free topology of e-mail networks.
\newblock {\em Physical Review E}, 66:035103, 2002.

\bibitem{faloutsos99powerlaw}
M.~Faloutsos, P.~Faloutsos, and C.~Faloutsos.
\newblock On power-law relationships of the internet topology.
\newblock In {\em {SIGCOMM}}, pages 251--262, 1999.

\bibitem{girvan-2002-99:community}
M.~Girvan and M.~E.~J. Newman.
\newblock Community structure in social and biological networks.
\newblock {\em Proc. Natl. Acad. Sci.}, 99:7821, 2002.

\bibitem{gokhale2008:hardware}
M.~Gokhale, J.~Cohen, A.~Yoo, W.~M. Miller, A.~Jacob, C.~Ulmer, and R.~Pearce.
\newblock Hardware technologies for high-performance data-intensive computing.
\newblock {\em Computer}, 41(4):60--68, 2008.

\bibitem{govindan00:heuristics}
R.~Govindan and H.~Tangmunarunkit.
\newblock Heuristics for internet map discovery.
\newblock In {\em {IEEE INFOCOM} 2000}, pages 1371--1380, Tel Aviv, Israel,
  March 2000.

\bibitem{hernandez2005:SLEPc}
V.~Hernandez, J.~E. Roman, and V.~Vidal.
\newblock Slepc: A scalable and flexible toolkit for the solution of eigenvalue
  problems.
\newblock {\em ACM Trans. Math. Softw.}, 31(3):351--362, 2005.

\bibitem{ipsen2005:analysis}
I.~Ipsen and R.~M. Wills.
\newblock Analysis and computation of google's pagerank.
\newblock In {\em Proc. 7th IMACS International Symposium on Iterative Methods
  in Scientific Computing}, 2005.

\bibitem{johnson1996:cliques}
D.~Johnson and M.~Tric.
\newblock Cliques, coloring and satisability.
\newblock In {\em Second DIMACS Implementation Challenge}. American
  Mathematical Society, Oct. 1996.

\bibitem{karypis95metis}
G.~Karypis and V.~Kumar.
\newblock {\em MeTis: Unstrctured Graph Partitioning and Sparse Matrix Ordering
  System, Version 2.0}, 1995.

\bibitem{krapivsky2000:connectivity}
P.~L. Krapivsky, S.~Redner, and F.~Leyvraz.
\newblock Connectivity of growing random networks.
\newblock {\em Physical Review Letters}, 85:4629, 2000.

\bibitem{lumsdaine07:_challenges_in_parallel_graph}
A.~Lumsdaine, D.~Gregor, B.~Hendrickson, and J.~Berry.
\newblock Challenges in parallel graph processing.
\newblock {\em Parallel Processing Letters}, 17(1):5--20, 2007 2007.

\bibitem{newman:2001b}
M.~E.~J. Newman.
\newblock From the cover: The structure of scientific collaboration networks.
\newblock {\em Proceedings of the National Academy of Sciences}, 98:404--409,
  2001.

\bibitem{newman:2001c}
M.~E.~J. Newman.
\newblock Scientific collaboration networks ii. shortest paths, weighted
  networks, and centrality.
\newblock {\em Phys. Rev. E}, 64:016132, 2001.

\bibitem{newman:2004b}
M.~E.~J. Newman.
\newblock Detecting community structure in networks.
\newblock {\em European Physical Journal B}, 38:321--330, May 2004.

\bibitem{newman:2004c}
M.~E.~J. Newman.
\newblock Fast algorithm for detecting community structure in networks.
\newblock {\em Phys. Rev. E}, 69(6):066133, June 2004.

\bibitem{newman:2004a}
M.~E.~J. Newman and M.~Girvan.
\newblock Finding and evaluating community structure in networks.
\newblock {\em Phys. Rev. E}, 69(2):026113, Feb. 2004.

\bibitem{Page99thepagerank}
L.~Page, S.~Brin, R.~Motwani, and T.~Winograd.
\newblock The pagerank citation ranking: Bringing order to the web, 1999.

\bibitem{satorras2001:dynamical}
R.~Pastor-Satorras, A.~Vazquez, and A.~Vespignani.
\newblock Dynamical and correlation properties of the internet.
\newblock {\em Physical Review Letters}, 87:258701, 2001.

\bibitem{redner1998:how}
S.~Redner.
\newblock How popular is your paper? an empirical study of the citation
  distribution.
\newblock {\em The European Physical Journal B}, 4:131, 1998.

\bibitem{shannon51:prediction}
C.~Shannon.
\newblock Prediction and entropy of printed english.
\newblock Technical report, The Bell System Technical Journal, 1951.

\bibitem{tong2008:Random}
H.~Tong, C.~Faloutsos, and J.-Y. Pan.
\newblock Random walk with restart: fast solutions and applications.
\newblock {\em Knowl. Inf. Syst.}, 14(3):327--346, 2008.

\bibitem{trilinos}
Trilinos.
\newblock The trilinos project.
\newblock {\tt http://trilinos.sandia.gov}.

\bibitem{tsallis2000:are}
C.~Tsallis and M.~P. de~Albuquerque.
\newblock Are citations of scientific papers a case of nonextensivity?
\newblock {\em The European Physical Journal B}, 13:777, 2000.

\bibitem{vazquez2001:statistics}
A.~Vazquez.
\newblock Statistics of citation networks, 2001.
\newblock cond-mat/0105031.

\bibitem{vazquez2002:large}
A.~Vazquez, R.~Pastor-Satorras, and A.~Vespignani.
\newblock Large-scale topological and dynamical properties of internet.
\newblock {\em Physical Review E}, 65:066130, 2002.

\bibitem{Watts98}
D.~J. Watts and S.~H. Strogatz.
\newblock Collective dynamics of |[lsquo]|small-world|[rsquo]| networks.
\newblock {\em Nature}, 393(6684):440--442, June 1998.

\bibitem{watts98:collective}
D.~J. Watts and S.~H. Strogatz.
\newblock Collective dynamics of small-world networks.
\newblock {\em Nature}, 393:440--442, 4 June 1998.

\bibitem{yoo2006:parallel}
A.~Yoo and K.~Henderson.
\newblock Parallel massive scale-free graph generators.
\newblock In {\em Proc. SC2006}, 2006.

\end{thebibliography}
\end{document}